\documentclass[twocolumn]{extreme_UV_slopes_z10}

\newcommand{\oiiiuv}{O~III] $\lambda 1666$}
\newcommand{\ciii}{C~III] $\lambda 1908$}
\newcommand{\civ}{C~IV $\lambda 1549$}
\newcommand{\heii}{He~II $\lambda 1640$}

\newcommand{\oii}{[O~II] $\lambda 3727$}
\newcommand{\niii}{N~III] $\lambda 1750$}
\newcommand{\niv}{N~IV] $\lambda 1488$}
\newcommand{\mgii}{Mg~II] $\lambda 2880$}
\newcommand{\nev}{[Ne~V] $\lambda 3426$}
\newcommand{\neiii}{[Ne~III] $\lambda 3868$}
\newcommand{\He}{H$\epsilon$ + [Ne~III] $\lambda 3970$}
\newcommand{\Hd}{H$\delta$ $\lambda 4102$}
\newcommand{\Hg}{H$\gamma$ $\lambda 4341$}
\newcommand{\Hb}{H$\beta$ $\lambda 4862$}
\newcommand{\oiiiA}{[O~III] $\lambda 4363$}
\newcommand{\oiiida}{[O~III] $\lambda 4959$}
\newcommand{\oiiidb}{[O~III] $\lambda 5007$}
\newcommand{\sii}{[S~II] $\lambda\lambda 6717,31$}
\usepackage{rotating,tabularx}
\usepackage{amsmath}

\shorttitle{Extreme UV-slopes at $z=10$}
\shortauthors{Donnan et al.}
\graphicspath{{./}{figures/}}

\begin{document}

\title{Very bright, very blue, and very red: JWST CAPERS analysis of highly luminous galaxies with extreme UV slopes at $\mathbf{z = 10}$}

\correspondingauthor{Callum T. Donnan}
\email{callum.donnan@noirlab.edu}

\author[0000-0002-7622-0208]{Callum T. Donnan}
\affiliation{NSF's National Optical-Infrared Astronomy Research Laboratory, 950 N. Cherry Ave., Tucson, AZ 85719, USA}

\author[0000-0001-5414-5131]{Mark Dickinson}
\affiliation{NSF's National Optical-Infrared Astronomy Research Laboratory, 950 N. Cherry Ave., Tucson, AZ 85719, USA}

\author[0000-0003-1282-7454]{Anthony J. Taylor}
\affiliation{Department of Astronomy, The University of Texas at Austin, Austin, TX, USA}

\author[0000-0002-7959-8783]{Pablo Arrabal Haro}
\altaffiliation{NASA Postdoctoral Fellow}
\affiliation{Astrophysics Science Division, NASA Goddard Space Flight Center, 8800 Greenbelt Rd, Greenbelt, MD 20771, USA}

\author[0000-0001-8519-1130]{Steven L. Finkelstein}
\affiliation{Department of Astronomy, The University of Texas at Austin, Austin, TX, USA}

\author[0000-0002-0827-9769]{Thomas M. Stanton}
\affiliation{Institute for Astronomy, University of Edinburgh, Royal Observatory, Edinburgh EH9 3HJ, UK}

\author[0000-0003-1187-4240]{Intae Jung}
\affiliation{Space Telescope Science Institute, 3700 San Martin Drive, Baltimore, MD 21218, USA}

\author[0000-0001-7503-8482]{Casey Papovich}
\affiliation{Department of Physics and Astronomy, Texas A\&M University, College Station, TX, 77843-4242 USA}
\affiliation{George P.\ and Cynthia Woods Mitchell Institute for Fundamental Physics and Astronomy, Texas A\&M University, College Station, TX, 77843-4242 USA}

\author[0000-0003-3596-8794]{Hollis B. Akins}
\affiliation{Department of Astronomy, The University of Texas at Austin, Austin, TX, USA}

\author[0000-0002-6610-2048]{Anton M. Koekemoer}
\affiliation{Space Telescope Science Institute, 3700 San Martin Drive, Baltimore, MD 21218, USA}

\author[0000-0003-4368-3326]{Derek J. McLeod}
\affiliation{Institute for Astronomy, University of Edinburgh, Royal Observatory, Edinburgh EH9 3HJ, UK}

\author[0000-0002-8951-4408]{Lorenzo Napolitano}
\affiliation{INAF - Osservatorio Astronomico di Roma, Via Frascati 33, 00078 Monte Porzio Catone, Rome, Italy}
\affiliation{Dipartimento di Fisica, Università di Roma Sapienza, Città Universitaria di Roma - Sapienza, Piazzale Aldo Moro, 2, 00185, Roma, Italy}

\author[0000-0001-5758-1000]{Ricardo O. Amor\'{i}n}
\affiliation{Instituto de Astrof\'{i}sica de Andaluc\'{i}a (CSIC), Apartado 3004, 18080 Granada, Spain}

\author[0000-0003-0629-8074]{Ryan Begley}
\affiliation{Institute for Astronomy, University of Edinburgh, Royal Observatory, Edinburgh EH9 3HJ, UK}

\author[0000-0002-4193-2539]{Denis Burgarella}
\affiliation{Aix Marseille Univ, CNRS, CNES, LAM Marseille, France}

\author[0000-0000-0000-0000]{Adam C. Carnall}
\affiliation{Institute for Astronomy, University of Edinburgh, Royal Observatory, Edinburgh EH9 3HJ, UK}

\author[0000-0002-0930-6466]{Caitlin M. Casey}
\affiliation{Department of Physics, University of California, Santa Barbara, CA 93106, USA}
\affiliation{Cosmic Dawn Center (DAWN), Niels Bohr Institute, University of Copenhagen, Jagtvej 128, K{\o}benhavn N, DK-2200, Denmark}

\author[0000-0003-2536-1614]{Antonello Calabr\`o}
\affiliation{INAF - Osservatorio Astronomico di Roma, Via Frascati 33, 00078 Monte Porzio Catone, Rome, Italy}

\author[0000-0002-3736-476X]{Fergus Cullen}
\affiliation{Institute for Astronomy, University of Edinburgh, Royal Observatory, Edinburgh EH9 3HJ, UK}

\author[0000-0002-1404-5950]{James S. Dunlop}
\affiliation{Institute for Astronomy, University of Edinburgh, Royal Observatory, Edinburgh EH9 3HJ, UK}

\author[0000-0001-7782-7071]{Richard S. Ellis}
\affiliation{Department of Physics and Astronomy, University College London, Gower Street, London WC1E 6BT, UK}

\author[0000-0003-0531-5450]{Vital Fern\'andez}
\affiliation{Michigan Institute for Data Science, University of Michigan, 500 Church Street, Ann Arbor, MI 48109, USA}

\author[0000-0002-7831-8751]{Mauro Giavalisco}
\affiliation{University of Massachusetts Amherst, 710 North Pleasant Street, Amherst, MA 01003-9305, USA}

\author[0000-0002-3301-3321]{Michaela Hirschmann}
\affiliation{Institute of Physics, Laboratory of Galaxy Evolution, Ecole Polytechnique Federale de Lausanne (EPFL), Observatoire de Sauverny, 1290 Versoix, Switzerland}

\author[0000-0003-3424-3230]{Weida Hu}
\affiliation{Department of Physics and Astronomy, Texas A\&M University, College Station, TX, 77843-4242 USA}
\affiliation{George P.\ and Cynthia Woods Mitchell Institute for Fundamental Physics and Astronomy, Texas A\&M University, College Station, TX, 77843-4242 USA}

\author[0000-0002-8096-2837]{Garth Illingworth}
\affiliation{Department of Astronomy and Astrophysics, UCO/Lick Observatory, University of California, Santa Cruz, CA 95064, USA}

\author[0000-0001-9187-3605]{Jeyhan S. Kartaltepe}
\affiliation{Laboratory for Multiwavelength Astrophysics, School of Physics and Astronomy, Rochester Institute of Technology, 84 Lomb Memorial Drive, Rochester, NY 14623, USA}

\author[0000-0002-8360-3880]{Dale D. Kocevski}
\affiliation{Department of Physics and Astronomy, Colby College, Waterville, ME 04901, USA}

\author[0000-0002-5588-9156]{Vasily Kokorev}
\affiliation{Department of Astronomy, The University of Texas at Austin, Austin, TX, USA}

\author[0000-0000-0000-0000]{Ho-Hin Leung}
\affiliation{Institute for Astronomy, University of Edinburgh, Royal Observatory, Edinburgh EH9 3HJ, UK}

\author[0000-0003-1581-7825]{Ray A. Lucas}
\affiliation{Space Telescope Science Institute, 3700 San Martin Drive, Baltimore, MD 21218, USA}

\author[0000-0003-4965-0402]{Alexa M.\ Morales}
\affiliation{Department of Astronomy, The University of Texas at Austin, Austin, TX, USA}

\author[0000-0000-0000-0000]{Ross McLure}
\affiliation{Institute for Astronomy, University of Edinburgh, Royal Observatory, Edinburgh EH9 3HJ, UK}

\author[0000-0001-8940-6768]{Laura Pentericci}
\affiliation{INAF - Osservatorio Astronomico di Roma, Via Frascati 33, 00078 Monte Porzio Catone, Rome, Italy}

\author[0000-0003-4528-5639]{Pablo G. P\'erez-Gonz\'alez}
\affiliation{Centro de Astrobiolog\'{\i}a (CAB), CSIC-INTA, Ctra. de Ajalvir km 4, Torrej\'on de Ardoz, E-28850, Madrid, Spain}

\author[0000-0002-6748-6821]{Rachel S. Somerville}
\affiliation{Center for Computational Astrophysics, Flatiron Institute, 162 5th Avenue, New York, NY, 10010, USA}

\author[0000-0000-0000-0000]{Struan Stevenson}
\affiliation{Institute for Astronomy, University of Edinburgh, Royal Observatory, Edinburgh EH9 3HJ, UK}

\author[0000-0002-1410-0470]{Jonathan R. Trump}
\affiliation{Department of Physics, 196 Auditorium Road, Unit 3046, University of Connecticut, Storrs, CT 06269, USA}

\author[0000-0003-3466-035X]{{L. Y. Aaron} {Yung}}
\affiliation{Space Telescope Science Institute, 3700 San Martin Drive, Baltimore, MD 21218, USA}

\author[0000-0002-7051-1100]{Jorge A. Zavala}
\affiliation{University of Massachusetts Amherst, 710 North Pleasant Street, Amherst, MA 01003-9305, USA}



\begin{abstract}

We present \textit{JWST}/NIRSpec PRISM observations of three luminous ($M_{\rm UV}<-20$) galaxies at $z\sim10$ observed with the CAPERS Cycle 3 program. These galaxies exhibit extreme UV slopes compared to typical galaxies at $z=10$. Of the three sources, two of them are a close pair (0.22\arcsec) of blue galaxies at $z=9.800\pm0.003$ and $z=9.808\pm0.002$ with UV slopes of $\beta=-2.87\pm0.15$ and $\beta=-2.46\pm0.10$ respectively, selected from PRIMER COSMOS NIRCam imaging. We perform spectrophotometric modeling of the galaxies which suggests extremely young stellar ages and a lack of dust attenuation. For the bluest galaxy, its UV slope also suggests significant Lyman continuum escape. 
In contrast, the third source (selected from CEERS NIRCam imaging) at $z=9.942\pm0.002$ exhibits a red UV slope with $\beta=-1.51\pm0.08$. We rule out the possibility of a strong nebular continuum due to the lack of a Balmer jump and find no evidence to support the presence of active galactic nucleus continuum due to a lack of strong UV emission lines and no broad component to H$\gamma$ or H$\beta$. Instead, it is most likely that the red UV slope is due to dust-reddening ($A_{\rm V}\simeq0.9$) implying a significant level of dust-obscured star-formation only $\simeq480\, \rm Myr$ after the Big Bang. Under standard assumptions for dust attenuation, EGS-25297 would be the most intrinsically UV-luminous galaxy ($M_{\mathrm{UV,corr}}\simeq -22.4^{+0.7}_{-1.1}$) yet spectroscopically confirmed at $z \sim 10$. This work highlights that luminous galaxies at $z\gtrsim10$ have a diversity of dust properties and that spectroscopy of these galaxies is essential to fully understand star-formation at $z\gtrsim10$. 

\end{abstract}



\section{Introduction} \label{sec:intro}
Understanding the formation and evolution of the first populations of galaxies in the universe is crucial for building a complete understanding of galaxy evolution. Prior to the launch of \textit{JWST}, galaxies had been detected out to $z\sim10$ from both space and ground-based telescopes \citep[e.g.,][]{ellis2013,mclure2013,oesch2014,oesch2018,bowler2014,bowler2015,bowler2020,finkelstein2015,mcleod2015,mcleod2016,bouwens2021,bouwens2022}.
However, the launch of \textit{JWST} has allowed the robust detection of galaxies at $z>10$ for the first time due to the highly sensitive and multi-filter imaging at $\lambda=1-5\, \mu$m. This has produced large samples of galaxy candidates at $z>10$ \citep[e.g.,][]{castellano2022,naidu2022,donnan2023a, donnan2023b,adams2022, finkelstein2022c,harikane2023a,franco2024}. These samples of galaxies have been used to measure the evolution of the galaxy luminosity function (LF) showing a slow, gradual evolution out to $z\simeq14$ \citep[e.g.,][]{finkelstein2024, perezgonzalez2023,2025arXiv250315594P, mcleod2023, adams2023, donnan2024,whitler2025}. This was in agreement with some pre-\textit{JWST} observations \citep[][]{mcleod2016} but was a slower evolution than many pre-\textit{JWST} theoretical models \citep[e.g.,][]{mason2015, tacchella2018, yung2019}, particularly at the bright end (e.g., $M_{\rm UV}<-20$).

There have been a number of suggested alterations to the physical properties of galaxies at $z\geq10$ to explain this evolution in the UV LF. For example, increases in the star-formation efficiency \citep{harikane2023a,dekel2023,somerville2025}, a lack of dust-attenuation \citep{ferrara2023,ferrara2024}, active galactic nuclei (AGN) \citep{pacucci2022,hegde2024}, and/or a more top-heavy initial mass function \citep{hutter2025} have all been suggested.

However, the UV LF at $z>10$ (and the galaxy stellar mass function) can be successfully modeled without any of these changes. A simple model where stellar ages decrease with increasing redshift is sufficient \citep{donnan2025} to explain the abundance of galaxies through decreasing the mass-to-light ratio at $z\geq10$. Similarly, bursty star formation can also lower the mass-to-light ratio required to reproduce the UV LF \citep{gelli2024,yung2024a,kokorev2025}. Although modeling the UV LF can be achieved with a simple model, breaking degeneracies between these different potential explanations and understanding the diversity of the galaxy population at $z\geq10$ represents a more significant challenge. Detailed spectroscopic observations of galaxies at $z\geq10$ are vital to fully understand this population.

Early \textit{JWST} NIRSpec observations have successfully confirmed the redshifts of a number of galaxies at $z\geq10$ \citep[e.g.,][]{curtislake2022, bunker2023, arrabalharo2023, carniani2024, castellano2024,napolitano2024, napolitano2025, kokorev2025,naidu2025} with an ever growing number of NIRSpec observations of highly luminous ($M_{\rm UV}<-20$) galaxies at $z>9$. These observations have revealed a diverse range of chemical properties with some galaxies at $z>9$ showing elevated nitrogen and sub-solar carbon abundances \citep[e.g.,][]{bunker2023,castellano2024} whereas others have weak or non-detected UV emission lines \citep[e.g.,][]{curtislake2022,carniani2024}. There is also evidence of AGN activity in some of these sources \citep[e.g.][]{napolitano2024,taylor2025}. 

An important property of high-$z$ galaxies is their rest-frame UV-continuum slope, $\beta$ (where $f_{\lambda} \propto \lambda^{\beta}$). The slope of the UV stellar continuum is primarily sensitive to the age, metallicity, and dust attenuation of a galaxy \citep[e.g.,][]{calzetti1994,schaerer2002,calabro2021} with bluer/steeper slopes observed in galaxies that are younger, lower metallicity, and have less dust attenuation. However, at $z>9$, the dominant property that affects the stellar UV slope is dust attenuation \citep{tacchella2022}. The UV slope is also sensitive to the contribution of nebular continuum \citep{cameron2024,katz2024} and therefore the fraction of Lyman continuum radiation that escapes a galaxy, with bluer slopes for increasing $f_{\rm esc}$ \citep{chisholm2022,mascia2025,papovich2025}. The UV slopes of galaxies have been measured with \textit{HST} at $z=6-9$ showing an average slope of $\beta\simeq-2$ \citep{dunlop2012b,finkelstein2012,bouwens2014} aligning with theoretical predictions \citep{dayal2012}. The UV slope of galaxies has also been observed to evolve with both redshift and UV magnitude, with $\beta$ becoming increasingly blue with increasing redshift and decreasing UV luminosity \citep{rogers2014,austin2024}. 

\textit{JWST} has allowed the UV-continuum slope to be measured at $z\geq10$ for the first time, showing an increase in steepness to $\beta \simeq-2.6$ at $z\simeq12$, indicating a transition in the galaxy population to one which lacks significant dust attenuation \citep[e.g.,][]{morales2023,topping2023,cullen2023,dottorini2024}. Alternatively, it has been suggested that supernova-produced dust grains cause less UV attenuation \citep{mckinney2025} at fixed $A_{\rm V}$, due to potentially grayer attenuation laws. However, the measurement of $\beta$ for individual galaxies has been difficult with NIRCam due to the limited number of data points from broad photometric filters, with these results mainly describing the average galaxy population at a given redshift and UV luminosity. Therefore, it remains unclear how much scatter there truly is in the UV-continuum slope at $z>9$, with deep NIRSpec observations required to measure $\beta$ more precisely for individual galaxies. There have been a number of extremely blue UV slope (e.g., $\beta<-2.4$) candidates identified from NIRCam imaging \citep[e.g.,][]{topping2022}. However, it remains unclear if there are any particularly red galaxies (e.g., $\beta>-1.6$) at $z>9$ which could imply significant dust attenuation. Spectroscopy is crucial to confirm their UV slopes and understand what drives such extreme (either red or blue) UV slopes only $\sim500\, \rm Myr$ after the Big Bang.

In this paper we present three galaxies observed with NIRSpec in the CANDELS-Area Prism Epoch of Reionization Survey (CAPERS; GO-6368, PI M.\ Dickinson) selected for their extreme UV slopes at $z=10$. Two of these sources have extremely blue UV slopes ($\beta<-2.4$) and the other is extremely red ($\beta>-1.6$). We analyze their spectra to understand the UV continuum and physical properties of these extreme sources at $z\simeq10$. 

This paper is structured as follows. In Section~\ref{sec:data} we introduce the objects analyzed here, and describe the photometric and spectroscopic data used in this work. In Section~\ref{sec:analysis} we describe the redshift fitting, spectral energy distribution (SED) fitting and emission line fitting. In Section~\ref{sec:discussion1} and Section~\ref{sec:discussion2} we discuss the nature of the galaxies analyzed in this work. Finally, in Section~\ref{sec:conclusion} we summarize our conclusions. Throughout we use magnitudes in the AB system \citep{oke1974,oke1983}, a \citet{Kroupa2001} initial mass function, and assume a standard cosmological model with $H_0=70$ km s$^{-1}$ Mpc$^{-1}$, $\Omega_m=0.3$ and $\Omega_{\Lambda}=0.7$.  

\section{Observational data} \label{sec:data}

\subsection{Selection of targets}\label{sec:selection}
CAPERS is a JWST Cycle 3 Treasury Program that is carrying out NIRSpec multi-object spectroscopy with the low-resolution ($R\simeq30-300$) PRISM disperser for several thousand galaxies out to the highest accessible redshifts. CAPERS targets are selected from multi-band \textit{JWST}/NIRCam imaging in three of the legacy CANDELS \citep{grogin2011,koekemoer2011} fields. The Cosmic Evolution Survey \citep[COSMOS,][]{Scoville2007} and Ultra-Deep Survey \citep[UDS,][]{lawrence2007} fields were observed with NIRCam in the Public Release IMaging for Extragalactic Research survey (PRIMER; Dunlop et al.\ in prep.), while the Extended Groth Strip (EGS) was observed in the Cosmic Evolution Early Release Science \citep[CEERS;][]{finkelstein2025} program. 

A primary goal of CAPERS is to obtain NIRSpec PRISM spectra of galaxies at $z\gtrsim9$. The selection of these candidates is based on multiple samples with different selection criteria, in some cases selected from different photometric catalogs. A full description of CAPERS target selection and observations will be presented in future publications. 
In this paper, we focus on three galaxies (two of which form a very close pair) with spectroscopic redshifts $z \approx 10$  that lie at extremes of the UV continuum slope distribution:  very red, and very blue. There are a number of $z\gtrsim10$ galaxies successfully confirmed by CAPERS \citep[e.g.][]{kokorev2025}, however the galaxies presented in this work have extreme UV slopes compared to the full sample at this redshift (and indeed more extreme than other published CAPERS galaxies) and therefore form the focus of this paper.

EGS-25297 was originally selected as a highly robust $z\simeq10$ candidate by \citet{mcleod2023} with a photometric redshift of $z_{\rm phot} = 10.0^{+0.3}_{-0.5}$ (ID = CEERS-2-6-z10-7927). It was subsequently also selected as a $z=10$ candidate by \citet{finkelstein2024} at $z_{\rm phot}=10.09^{+0.18}_{-0.39}$ (ID=98518); in the Astrodeep catalog of \citet{merlin2024} (ID=17766, $z_{\rm phot} = 10.025$); and as a ``Little Red Dot'' (LRD; see Section~\ref{sec:25297_agn}) candidate by \cite{kocevski2025} (ID=23931, $z_{\rm phot} = 10.21$) due to its red photometric colors and ``v-shaped'' SED.

\begin{figure}[ht]
\centering
\includegraphics[width=\linewidth]
{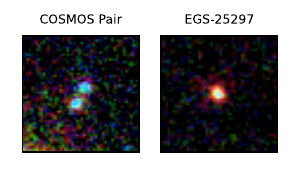}
\caption{RGB color images of COSMOS-300001 (left panel, upper-right source) and COSMOS-52185 (left panel, lower-left source) and EGS-25297 (right) using the \textit{JWST}/NIRCam F150W, F200W and F444W filters. Each cutout is $1.5\times1.5$ arcsec. The color images illustrate the blue rest-frame UV colors of the COSMOS pair and the red rest-frame UV color of EGS-25297.}
\label{fig:stamps}
\end{figure}

COSMOS-300001 and COSMOS-52185\footnote{The CAPERS ID for COSMOS-300001 was assigned by hand because the source catalog used for target selection incorrectly deblended that component of the pair.} are a pair of similarly bright ($m(\rm F277W) \approx 27.55$), compact objects separated by 0.22\arcsec. They  were identified as robust $z=10$ candidates in \citet{donnan2024} with photometric redshifts $z_{\rm phot}=10.0^{+0.37}_{-0.41}$ and $z_{\rm phot}=10.0^{+0.60}_{-0.42}$ respectively. The pair was identified as one source in the Astrodeep catalog of \citet{merlin2024} (ID=7462, $z_{\rm phot} = 10.05$), and also as one source, also with $z_{\rm phot} = 10.05$, in a sample of very high redshift candidates from COSMOS-Web (M.\ Franco, priv.\ comm., extending an earlier sample from \citealp{franco2024}). In Fig.~\ref{fig:stamps} we show color images of the sources analyzed in this work illustrating the contrasting blue colors for the COSMOS pair with the red color of EGS-25297.
The sample is described in Table~\ref{tab:sample}.

\begin{deluxetable*}{lccccc}
\label{tab:sample}
\caption{Properties of the $z=10$ galaxies in this work}
\tablehead{ID & RA & DEC & $z_{\rm spec}$ & $M_{\mathrm{UV}}$ & $\beta$ \\ 
& /deg & /deg & & /AB mag &  }

\startdata
  COSMOS-300001 & 150.1165406 & 2.1971091 & $9.800\pm0.003$ & $-20.3$5 & $-2.87\pm0.15$  \\
    COSMOS-52185 & 150.1165670 & 2.1970555 & $9.808\pm0.002$ & $-20.33$ & $-2.43\pm0.10$  \\
    EGS-25297 & 214.8171062 & 52.7483418 & $9.942\pm0.002$ & $-20.06$ & $-1.51\pm0.08$  \\
\enddata
\end{deluxetable*}

\subsection{Photometry}\label{sec:photometry}
All three CAPERS survey fields have \textit{JWST}/NIRCam imaging in the F090W, F115W, F150W, F200W, F277W, F356W, F410M, and F444W filters, as well as the \textit{HST}/ACS F606W and F814W filters. 
The photometry for EGS-25297 was measured using the default catalog used for CAPERS selection.  This photometric technique is similar to that in \citet{finkelstein2024}, and will be described in full in Finkelstein (in prep.). 
As the default catalog does not correctly split the PRIMER-COSMOS pair, for these sources we measure photometry following a procedure described in detail in \citet{donnan2024}; we provide a short description here. Sources are detected using \textsc{Source Extractor} \citep{bertin1996} in the F150W, F200W, and F277W images. The photometry is then measured using 0.3-arcsec diameter apertures on images for which the point spread functions (PSFs) have been homogenized to match that of the F444W filter (which has a PSF full width at half maximum [FWHM] of 0.145-arcsec). It was then corrected to total from the flux enclosed for the F444W PSF curve-of-growth. For both systems, we use the NIRCam photometry in combination with the NIRSpec spectrum to assist in modelling the SED of the sources as well as for correcting NIRSpec slit losses. 

Although the PRIMER and CEERS surveys contain some level of MIRI imaging that overlaps with the NIRCam coverage, they do not have coverage for these sources. The COSMOS pair is covered by F770W from the COSMOS-Web program \citep{caset2023b} and F2100W from COSMOS-3D \citep{kakiichi2024}, but the galaxies are undetected in both filters.

\subsection{Spectroscopy}
CAPERS NIRSpec observations are split into 21 pointings (7 per survey field). Each pointing is observed using three NIRSpec Multi-Shutter Assembly (MSA) configurations. The highest priority sources are observed in all 3 configurations per pointing to maximize the observing time. This is the case for EGS-25297 with a total observing time of $17069\, \rm s$ ($4.74\, \rm hr$). 
However, although COSMOS-300001 and COSMOS-52185 are both classified as high priority targets, due to their close proximity to each other and the  aperture position for the scheduled NIRSpec observations, the $4.74\, \rm hr$ observing time was split between the 2 sources. COSMOS-300001 was observed in a single configuration with an observing time of $5689.67\, \rm s$ ($1.58\, \rm hr$) and COSMOS-52185 was observed in 2 configurations with an observing time of $11379 \, \rm s$ ($3.16\, \rm hr$). 

We reduce the NIRSpec data using version 1.17.1 of the \textit{JWST} Calibration Pipeline \citep{bushouse2025} with CRDS version \texttt{1350.pmap} following the general method described in \citet{arrabalharo2023b}. We do this with a number of alterations. 
Firstly, we use the \texttt{clean\_flicker\_noise} option to remove the effect of $1/f$ noise in the \texttt{calwebb\_detector1} stage. In \texttt{calwebb\_spec2} we also use a modified flat-field . After running the \texttt{calwebb\_spec3} stage the product was a reduced 2D spectrum. We then extract the 1D spectra for each source using an optimal extraction \citep{horne1986}. The 2D and 1D spectra for the 3 sources are shown in Fig.~\ref{fig:spectra}.

Finally, we correct the spectra to account for NIRSpec slit losses and flux calibration uncertainties. To do this, we first derive the synthetic photometry from the NIRSpec spectra for all the relevant NIRCam filters. We then measure the relative offset in each filter and perform a linear fit as a function of wavelength which provides a good fit to the photometry offset. We then apply this wavelength dependent correction to the NIRSpec spectra.
 
The Dawn JWST Archive \footnote{https://dawn-cph.github.io/dja/} (DJA) v4 NIRSpec data release extends the wavelength range for the PRISM spectrum of EGS-25297 to $5.5\,\mu$m. We use the DJA spectrum to measure emission line fluxes for H$\beta$, \oiiida \, and \oiiidb\ (see Appendix~\ref{sec:apdx_A}). 

\section{Data analysis} \label{sec:analysis}

\subsection{Spectroscopic redshifts}

All three objects show very strong Ly$\alpha$ breaks near $\lambda_{\rm obs} \approx  1.3\,\mu$m.  For EGS-25297, the red end of the spectrum shows very strong emission lines from [O~II], [Ne~III], H$\delta$ and blended H$\gamma$+\oiiiA. These lines are weaker for COSMOS-300001 and COSMOS-52185, and thanks to their slightly lower redshifts  H$\beta$ is also detected. Possible detections of UV rest-frame emission lines are discussed below.

The measurement of spectroscopic redshifts in CAPERS is being undertaken by multiple redshift fitting codes as well as visual inspection. In this paper we provide a redshift measurement using the SED fitting code \textsc{bagpipes} \citep{carnall2018}. Due to the detection of a number of emission lines, we are able to obtain a more precise redshift than is possible with only a Ly$\alpha$ break \citep[e.g.,][]{curtislake2022, arrabalharo2023b}. For COSMOS-300001 we measure $z_{\rm spec}=9.800\pm0.003$ and for COSMOS-52185 we measure $z_{\rm spec}=9.808\pm0.002$. For EGS-25297 we measure $z_{\rm spec}=9.942\pm0.002$.

\begin{figure*}[ht]
\centering
\includegraphics[width=\linewidth]
{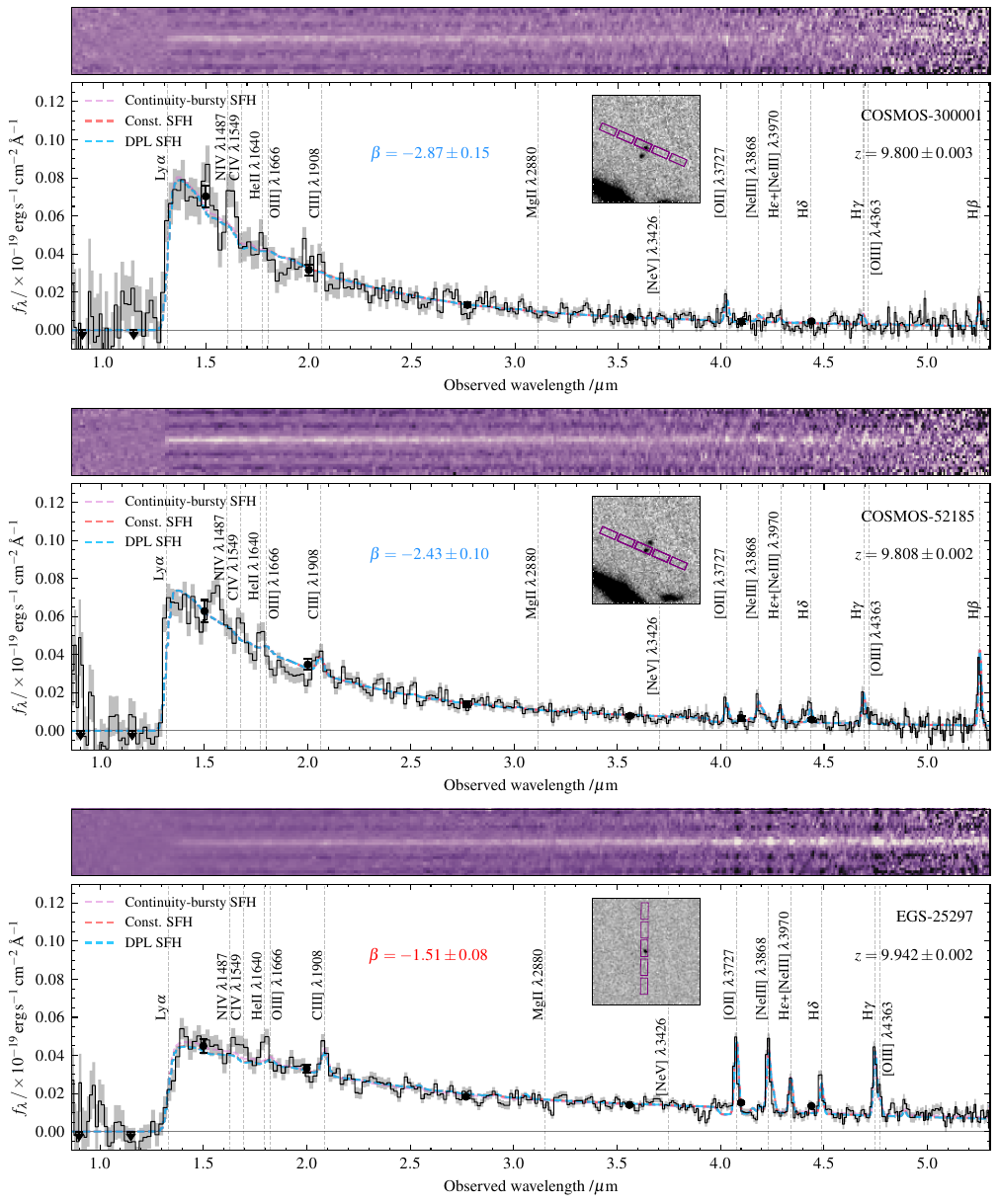}
\caption{The 2D (top) and 1D (bottom) optimally extracted spectra of the three sources presented in this work, COSMOS-300001 (top), COSMOS-52185 (middle) and EGS-25297 (bottom). The black data points show the photometry for each galaxy. For each spectrum we highlight the position of common emission lines (dashed gray line) and plot the best-fitting \textsc{bagpipes} SED model assuming a constant (red), double power-law (blue) and Continuity-bursty non-parametric (purple) star-formation history prior. We show $3\times3$ arcsec cutout images in the F150W filter for each source with the NIRSpec slit overlayed (inset).}
\label{fig:spectra}
\end{figure*}

\subsection{SED fitting}
To measure a number of physical properties of the galaxies, we further use the SED fitting code \textsc{BAGPIPES} to model the spectra and photometry of each galaxy simultaneously (a full description of how this works is available in \citealp{carnall2019}). The SED modeling with \textsc{bagpipes} accounts for the variable spectral resolution of the NIRSpec PRISM spectra. We use \citet{bruzual2003} stellar population models with the MILES stellar library \citep{Falcon-Barroso2011}. We assume a \citet{calzetti2000} dust-attenuation law with a uniform prior on the V-band attenuation over the range $A_{\rm V}=0-3$. We also assume a log-uniform prior on the stellar mass where $\log(M_{\star}/\rm M_{\odot}) = 5.0-13.0$, a log-uniform prior on metallicity where $Z/ \mathrm{Z_{\odot}} = 0 - 2.5$ and a log-uniform prior on the ionization parameter where $-4<\log U<0$. We assume a \citet{Kroupa2001} initial mass function in this fitting. The inferred star-formation history (SFH) of a galaxy through SED fitting is often sensitive to the assumed parametric form of the SFH. Here we test a number of different parametric and non-parametric forms for the SFH when fitting the SED. Firstly, we adopt a constant SFH prior where we place a log-uniform prior on the stellar age between $1\, \rm Myr$ and $10\, \rm Gyr$. Secondly, we adopt a double power-law (DPL) prior which parametrizes the SFH with rising and falling power-laws. Here we impose a log-uniform prior on each slope ($d\log(\mathrm{SFR})/d\log t$) between 0.01 and 1000. Finally, we adopt a non-parametric prior on the SFH. We assume a ``continuity-bursty'' prior as defined by \citet{tacchella2022} where we allow the SFH to change over 8 bins in cosmic time. The first 4 bins are spaced logarithmically between $z=20$ and $100\, \rm Myr$ with the final 4 bins defined between $30-100 \, \rm Myr$, $10-30 \, \rm Myr$, $3-10 \, \rm Myr$ and  $0-3 \, \rm Myr$. 

\begin{deluxetable*}{lcccccc}
\label{tab:SED_properties}
\caption{Properties derived from SED fitting of the $z=10$ galaxies in this work}
\tablehead{ID & $\log(M_{\star}/ \mathrm{M_{\odot}})$ & $t_{\star}$ & $A_{\rm V}$ & $\rm SFR_{10}$ & $\log(U)$ & $Z$ \\ 
 & & /Myr & /mag & /$\rm M_{\odot}\, yr^{-1}$ & & /$\rm Z_{\odot}$ }

\startdata
  COSMOS-300001 & $7.81^{+0.07}_{-0.05}$ & $3.53^{+1.22}_{-2.52}$ & $0.02^{+0.03}_{-0.01}$ & $7^{+3}_{-2}$ & $-1.3^{+0.2}_{-0.2}$ & $0.9^{+0.3}_{-0.3}$ \\
    COSMOS-52185 & $7.90^{+0.05}_{-0.05}$ & $3.36^{+0.71}_{-1.25}$ & $0.07^{+0.05}_{-0.03}$ & $9^{+3}_{-2}$ & $-1.2^{+0.3}_{-0.3}$ & $0.13^{+0.02}_{-0.02}$\\
    EGS-25297 & $8.95^{+0.04}_{-0.05}$ & $2.63^{+0.23}_{-0.30}$ & $0.92^{+0.42}_{-0.27}$ & $117^{+21}_{-18}$ & $-1.28^{+0.07}_{-0.08}$ & $0.200^{+0.004}_{-0.004}$\\
\enddata
\tablecomments{The final six columns show the results of SED fitting using \textsc{bagpipes} (assuming a double-power law parametric form for the star-formation history) showing the stellar mass, mass-weighted stellar age, V-band dust attenuation, star-formation rate over the last $10\, \rm Myr$, ionization parameter and metallicity.}
\end{deluxetable*}

\begin{figure*}[ht]
\centering
\includegraphics[width=\linewidth]
{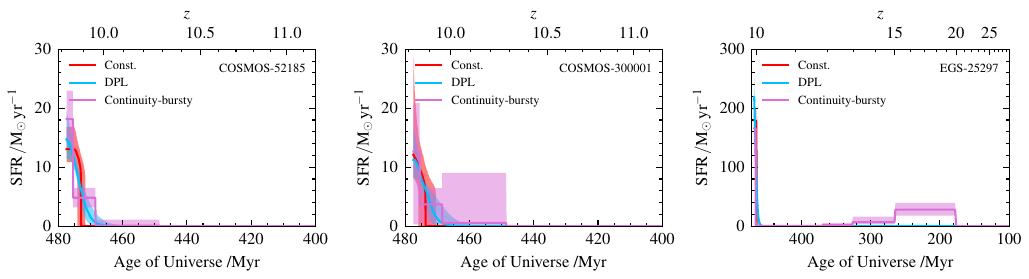}
\caption{The inferred star-formation histories (SFH) from the best-fitting \textsc{bagpipes} SED model to COSMOS-52185 (left), COSMOS-300001 (middle) and EGS-25297 (right). We show the inferred SFH assuming three different priors for each galaxy with a constant prior in red, a double power-law in blue and the ``continuity-bursty'' non-parametric prior \citep[][]{tacchella2022} in purple. }
\label{fig:SFH}
\end{figure*}

In Fig~\ref{fig:spectra} we show the best-fitting SED templates and in Fig~\ref{fig:SFH} we show the resulting star-formation histories. For the COSMOS pair we find a young burst of star-formation of $\sim10-15\, \rm M_{\odot}\, yr^{-1}$ with a mass-weighted age (the time since the galaxy was half its present stellar mass) of $t_{\star}\simeq3.4\, \rm Myr$ for both sources. We also find low stellar masses of $\log(M_{\star}/ \rm M_{\odot})\simeq7.9$ for both galaxies. This result is consistent for all three SFH priors which all show essentially indistinguishable results for both the SFH and resultant SED model. 
For EGS-25297, when assuming DPL or constant star formation, we find SFHs similar to those derived for the COSMOS pair galaxies, with a mass-weighted age of $t_{\star}=2.63^{+0.23}_{-0.30}\, \rm Myr$.
However, we find a stellar mass of $\log(M_{\star}/ \rm M_{\odot})=8.95^{+0.04}_{-0.05}$ which is $\sim1\, \rm dex$ greater than each of the COSMOS galaxies. Also, in contrast to the COSMOS pair, we obtain a different result when assuming a non-parametric ``continuity-bursty'' prior, with an episode of early star formation occurring at $z\simeq20$, resulting in a stellar mass  of $\log(M_{\star}/ \rm M_{\odot})=9.60^{+0.09}_{-0.12}$. The implications of this SFH are discussed in detail in Section~\ref{sec:25297_SFH}. We report stellar masses, mass-weighted stellar ages and V-band dust attenuation for all three galaxies in Table~\ref{tab:SED_properties}, assuming the DPL SFH prior.  

\subsection{Galaxy sizes}
To determine the physical sizes of the galaxies in this work, we use the \textsc{galfit} software \citep{peng2002} to model the 2-dimensional profile of the galaxies in different NIRCam filters. We fit a S\'{e}rsic profile with index $n=1$ and find that all three galaxies are consistent with being point-sources, with no evidence of extended emission. We place upper limits on the angular size as 0.04-arcsec in F200W and 0.08-arcsec in F444W. These correspond to upper limits on the physical sizes of $\lesssim170\, \rm pc$ and $\lesssim340\, \rm pc$ (F200W and F444W, respectively) for each galaxy. The FWHM of the F200W and F444W filters are 0.066-arcsec and 0.145-arcsec respectively. The pixel scale of the imaging is 0.03-arcsec.

\subsection{Emission lines}
\label{sec:emission_lines}
In Fig.~\ref{fig:spectra} we highlight the positions of typically strong rest-frame UV and optical galaxy emission lines as shown by the gray dashed lines. We initially measure the continuum of the spectra by adopting the best-fitting \textsc{bagpipes} model where we mask any emission lines. To initially assess which lines have been detected we perform a direct integration of the continuum-subtracted spectra at the expected wavelengths of these lines in a window of $4\sigma_{\rm R}(\lambda)$ where $\sigma_{\rm R}(\lambda)$ is the expected standard deviation of a Gaussian emission line determined by the variable spectral resolution of NIRSpec PRISM which increases from $R\simeq30$ to $R\simeq300$ as a function of wavelength. For emission lines with a $>3\sigma$ significance from this direct integration, we then determine the best-fitting model to each emission line. To do this we use a Levenberg-Marquardt method to fit a Gaussian profile at the position of each emission line. We estimate the uncertainty of the line fluxes by randomly perturbing the observed fluxes assuming they follow a Gaussian distribution with a standard deviation equal to the flux uncertainty. We repeat this 1000 times to create a distribution of line flux measurements and take the standard deviation of this distribution as the uncertainty on the line flux. The continuum measurement from the \textsc{bagpipes} model is used in the $\rm EW_0$ measurement.

Although the \Hg\, and \oiiiA\, emission lines are partially blended, the spectral resolution of the PRISM at its reddest wavelengths, in conjunction with a precise redshift from other emission lines, is sufficient to enable us to de-blend them in COSMOS-52185 and EGS-25927.  (No significant emission from \Hg+\oiiiA\, is detected for COSMOS-300001.) We therefore perform a 2-component Gaussian fit to extract the emission line fluxes for each line separately where each line shares the same width. In Table~\ref{tab:line_measurements} we report the measured line fluxes, rest-frame equivalent widths (EWs) and their associated uncertainties for a number of rest-frame UV and optical emission lines.

\begin{deluxetable*}{lcccccc}
\label{tab:line_measurements}
\caption{Emission line measurements for the $z\approx10$ galaxies in this work}
\tablehead{ & COSMOS-300001 &  & COSMOS-52185 &  & EGS-25297 &  \\
         Emission line & Flux  & EW$_0$  & Flux & EW$_0$  & Flux  & EW$_0$\\
          & /$10^{-19}\, \rm erg \, s^{-1}\, cm^{-2}$ & /$\rm \AA$ & /$10^{-19}\, \rm erg \, s^{-1}\, cm^{-2}$ & /$\rm \AA$ & /$10^{-19}\, \rm erg \, s^{-1}\, cm^{-2}$ & /$\rm \AA$ \\}

\startdata
  \niv & $<10.8$ & $<18.4$ & $<7.10$ & $<12.2 $ & $<5.86$ & $<13.6$ \\
  \civ & $<9.71$ & $<18.9$ & $<6.58$ & $<12.3$ & $<5.79$ & $<14.1$ \\
  \heii $+$ \oiiiuv  & $<8.41$ & $<18.8$ & $<5.74$ & $<12.2$ & $6.86\pm3.11$ & $17.8\pm8.1$ \\
  \ciii & $<6.08$ & $<18.7$ & $<4.22$ & $<12.8$ & $7.40\pm3.44$ & $22.6\pm10.5$ \\
  \mgii & $<1.99$ & $<18.9$ & $<1.71$ & $<14.4$ & $<1.50$ & $<8.2$ \\
  \oii & $2.64\pm0.71$ & $55.7\pm15.1$ & $2.20\pm0.60$ & $42.4\pm11.5$ & $7.79\pm0.52$ & $77.4\pm5.2$ \\
  \nev & $<1.53$ & $<23.0$ & $<1.32$ & $<16.0$ & $<1.19$ & $<7.83$  \\
  \neiii & $<1.82$ & $<36.8$ & $<1.31$ & $<20.6$ & $8.91\pm0.61$ & $71.5\pm4.9$ \\
  \He   & $<1.45$ & $<32.5$ & $<1.32$ & $<25.9$ & $3.44\pm0.49$ & $32.2\pm4.6$ \\
  \Hd & $<1.44$ & $<35.1$ & $3.07\pm1.07$ & $64.6\pm22.5$ & $4.17\pm0.61$ & $40.7\pm5.9$ \\
  \Hg & $<1.54$ & $<45.2$ & $2.26\pm0.56$ & $55.8\pm13.7$ & $6.79\pm0.75$ & $75.0\pm8.3$ \\
  \oiiiA & - & - & $0.99\pm0.45$ & $24.3\pm11.1$ & $2.68\pm0.50$ &$29.6\pm5.5$ \\
  \Hb & $2.51\pm0.77$ & $104\pm32$ & $7.68\pm2.02$ & $255\pm67$ & $17.7\pm1.4$ & $232\pm18$ \\
  \oiiida & $-$ & $-$ & $-$ & $-$ & $38.9\pm2.6$ & $-$ \\
  \oiiidb & $-$ & $-$ & $-$ & $-$ & $117\pm8$ & $-$ \\
\enddata
\tablecomments{Upper limits are shown as $2\sigma$ determined through direct integration of the spectra. The \Hb\, \oiiida\, and \oiiidb\, fluxes for EGS-25297 are measured using the extended wavelength reduction from the DJA.}
\end{deluxetable*}

\subsection{UV properties}
\begin{figure}[ht]
\centering
\includegraphics[width=\columnwidth]
{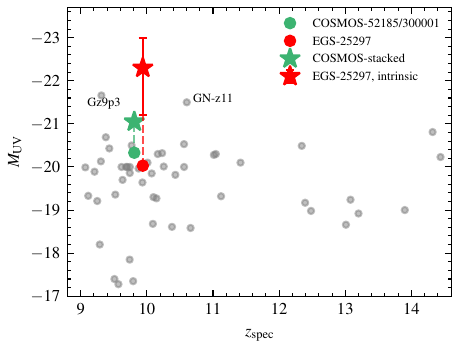}
\caption{A comparison of the UV magnitude and redshift of COSMOS-52185, COSMOS-300001 individually (green point) and stacked (green star), and EGS-25297 (red point) to the total sample of spectroscopically confirmed galaxies at $z>9$ in the literature. The red star shows the intrinsic UV magnitude of EGS-25297 after correction for dust attenuation using standard assumptions. The gray data-points show the observed values for other spectroscopically confirmed galaxies at $z>9$ \citep{yanagisawa2024,williams2023b,witstok2025, wang2023,tang2024,stiavelli2023,schaerer2024,runnholm2025,robertsborsani2023,napolitano2025,Hsaio2024, hainline2024, goulding2023,fujimoto2023,deugenio2024,curtislake2022,curti2024,castellano2024,carniani2024,bunker2023b,boyett2024,arrabalharo2023,arrabalharo2023b,kokorev2025,taylor2025,naidu2025}.}
\label{fig:Muv_z}
\end{figure}

We determine the UV absolute magnitude of each galaxy by directly integrating the photometry-corrected spectra for each component across a top-hat function centered at $\lambda_{\rm rest}=1500\, \rm \AA$ with a width of $100\, \rm \AA$. For EGS-25297 we measure $M_{\rm UV}=-20.0$. For both COSMOS-52185 and COSMOS-300001 we find similar UV magnitudes of $M_{\rm UV}=-20.3$ and if taken together, the COSMOS pair has $M_{\rm UV}=-21.05$. For EGS-25297 we measure a UV absolute magnitude of $M_{\rm UV}=-20.0$. These magnitudes are shown in context with other spectroscopically confirmed galaxies at $z>9$ in Fig.~\ref{fig:Muv_z}.

We also use the spectrum of each source to estimate its UV continuum slope, $\beta$.
We estimate $\beta$ by fitting a power-law to the photometry-corrected spectrum of the form $f_{\lambda}\propto \lambda^{\beta}$, over the wavelength range $\lambda_{\rm rest}=1350-2600 \,\mathrm{ \AA }$ to avoid the effect of damped Ly$\alpha$ absorption \citep{heintz2025}. For this calculation we mask the flux at the position of known emission lines (\niv\,, \heii\,, \civ\, \niii\, and \ciii). We determine the uncertainty on $\beta$ by scattering the flux at each pixel assuming Gaussian noise with a standard deviation equal to the flux uncertainty. The UV slope is then re-measured with this process then repeated 1000 times. For EGS-25297 we measure a UV-slope of $\beta=-1.51\pm0.08$, making it the reddest spectroscopically confirmed galaxy at $z>9$ to date. For COSMOS-52185 and COSMOS-300001 we find that both spectra exhibit blue UV-slopes with $\beta=-2.43\pm0.09$ and $\beta=-2.87\pm0.15$ respectively. 
For further verification, we also measure the UV-slope using photometry alone, using the technique described in \citet{cullen2023} and find consistent results (within $\Delta \beta\pm0.1$). Measuring the UV slope from the best fit SED models (from our spectrophotometric fits) yields redder UV slopes by $\Delta \beta \simeq0.4$. However, there is a bias towards redder UV slopes when measured from the best fitting SED model \citep{rogers2013} and this has been shown to be less accurate than fitting a power-law directly to the photometry \citep{morales2025}. 

\begin{figure*}[ht]
\centering
\includegraphics[width=\linewidth]
{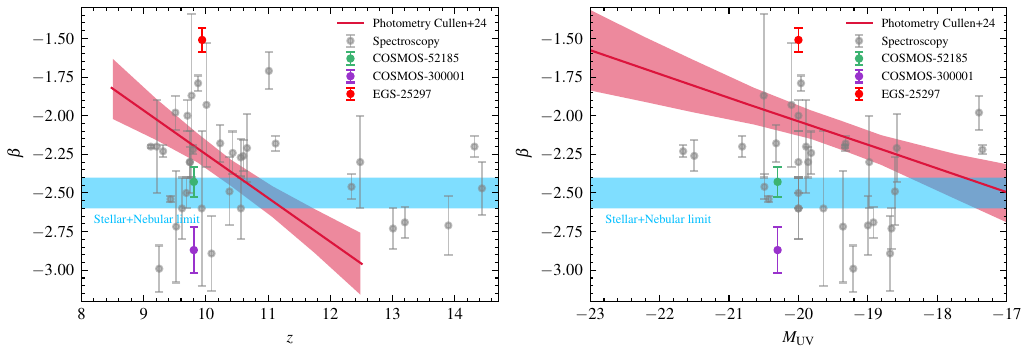}
\caption{The $\beta - z$ relation (left) and the $\beta - M_{\rm UV}$ (right) from \citet{cullen2023} (red line) in comparison to COSMOS-52185 (green), COSMOS-300001 (purple) and EGS-25297 (red). The gray data-points show other measurements from spectroscopically confirmed galaxies at $z>9$ (same references as Fig.~\ref{fig:Muv_z}). The blue horizontal region shows the steepest UV-slopes possible assuming a contribution from both stellar and nebular continua assuming no escape of Lyman continuum.}
\label{fig:beta_relations}
\end{figure*}

To place these measurements into context within the known galaxy population at $z>9$ we compare to previous photometric \citep[e.g.,][]{topping2023,morales2023,cullen2023} and spectroscopic observations. This is shown in Fig.~\ref{fig:beta_relations}. This demonstrates that these three galaxies represent extremes for the bright ($M_{\rm UV}<-20$) galaxy population at $z=10$. EGS-25297 has a significantly redder UV-slope than any other similar source in the literature. We discuss the physical interpretation of this measurement in Section~\ref{sec:25297_beta}. Conversely, both COSMOS-52185 and COSMOS-300001 have significantly bluer UV-slopes than are typical for galaxies at their redshifts and UV magnitude. In particular, COSMOS-300001 is one of the bluest sources found to date. We discuss the implications of such blue UV slopes in Section~\ref{sec:cosmos_beta}.

\subsection{Lyman~$\alpha$ damping}
\label{sec:ly_break}
The shape of the continuum break near the wavelength of Ly$\alpha$ is sensitive to the intergalactic and circumgalactic neutral hydrogen environment of galaxies. These three sources exhibit significantly different break morphologies (Fig.~\ref{fig:spectra}).  COSMOS-300001 and COSMOS-52185 show sharp breaks, with flux dropping to zero over a few pixels, not much greater than the spectral resolution of the NIRSpec PRISM. 
We model the Ly$\alpha$ damping wing following the method described in Section 3.7 of \citet{Hsaio2024}.  Specifically, we calculate the IGM Ly$\alpha$ damping wing optical depth using Equation~3 from \citet{Dijkstra2014}, which depends on two key parameters: the volume-averaged neutral hydrogen fraction ($X_{\mathrm{HI}}$) and the size of the ionized bubble ($R_{\mathrm{HII}}$) surrounding the source. Additionally, we account for the potential contribution from a damped Ly$\alpha$ absorber (DLA) along the line of sight, adopting the modeling approach of \citet{heintz2024}, where the DLA is characterized by its neutral hydrogen column density ($N_{\mathrm{HI}}$). We find $X_{\mathrm{HI}} > 0.55$ for both sources, with $\log(N_{\mathrm{HI}}/\mathrm{cm}^{-2}) \approx 17$ for COSMOS-300001, and negligible $n_{\mathrm{HI}}$ for COSMOS-52185.
In contrast, EGS-25297 has a much more gradual break, with best-fit parameters $X_{\mathrm{HI}} \approx 0.8$ and $\log(n_{\mathrm{HI}}/\mathrm{cm}^{-2}) = 21.9\pm0.3$.  This high neutral column density is inconsistent with normal IGM damping, and suggests the presence of a neutral gas reservoir local to the galaxy \citep[e.g.,][]{heintz2024}.

\section{A merger induced star-burst at $\lowercase{z}=9.8$?} \label{sec:discussion1}
\subsection{The nature of star-formation in the COSMOS pair}
COSMOS-300001 and COSMOS-52185 are a very close pair of nearly equally bright objects. If we were to treat them as a single galaxy\footnote{As noted in \S\ref{sec:data}, some source catalogs of the PRIMER COSMOS field detect the pair as a single object.} it would be the third brightest known at $z > 9$, at $M_{\rm UV}=-21.05$, with only Gz9p3 \citep{boyett2024} and GN-z11 \citep{bunker2023b} having brighter UV magnitudes as demonstrated in Fig.~\ref{fig:Muv_z}. Although \textit{JWST} has been able to spectroscopically confirm redshifts and measure properties of similarly luminous galaxies at $z\simeq10$ \citep[e.g.,][]{arrabalharo2023,napolitano2025,kokorev2025}, this COSMOS pair shows some notable differences. Firstly, from SED fitting the inferred stellar masses of COSMOS-300001 and COSMOS-52185 are significantly lower than those observed at similar UV luminosity and redshift by $\sim0.5\, \rm dex$ (although it is important to note that galaxy stellar masses are poorly constrained without longer wavelength observations \citep[][]{Papovich2023}). This implies a smaller mass-to-light ratio than has been typically observed in other similarly luminous galaxies \citep[e.g.][]{robertsborsani2024,kokorev2025}. Our best-fitting SED model from \textsc{bagpipes} implies that this is due to a strong burst in star-formation with a very young stellar population dominated by O and B type stars.

This naturally therefore raises the question of what is driving this intense burst of star-formation and extremely blue UV slope, and if these are connected to the fact they have close physical proximity to each other and may be about to merge. It has frequently been postulated that mergers could drive bursts of star-formation due to tidal forces from the galaxy that shock and compress the gas. This has been observed in simulations \citep{kim2009,saitoh2009} however observationally there has been limited evidence to support merger-induced star-bursts \citep{silva2018,pearson2019} at high-redshift. 

The transverse separation of the two galaxies is $\simeq0.93\, \rm kpc$ and their redshift separation $\Delta z$ corresponds to a velocity difference of $\Delta v=220\pm100\, \rm km\,s^{-1}$; both measurements may be underestimates of the true three dimensional quantities due to projection effects. The dynamical timescale corresponding to an interaction with this relative velocity and physical separation would be of the order of $\simeq4$ Myr.
If this system will undergo a merger then its true kinematic state would likely be more complex than the simple approximation obtained here from the projected separation and the line-of-sight velocity. Both galaxies still appear relatively distinct, although at such a large redshift cosmological dimming surely inhibits detection of low surface brightness features such as tidal debris and bridges. It seems likely that any merger may be in its early stages, possibly with just the initial passage having taken place. Hence this dynamical timescale of $\simeq4$ Myr may be of the same order of magnitude as the time elapsed since the initial passage, and it is interesting that this compares well to the inferred mass-weighted burst stellar ages of $3.53^{+1.22}_{-2.52}\, \rm Myr$ and $3.36^{+0.71}_{-1.25}\, \rm Myr$ in Table~\ref{tab:SED_properties}, as well those ages being fairly co-eval, thereby supporting a scenario where a merger or interaction may have played a role in triggering the star formation in this galaxy pair.

The close angular separation and relatively large (albeit uncertain) velocity difference between COSMOS-52185 and -300001, as well as their luminosities and stellar masses, suggest that they are two galaxies that may merge.  If, instead, we consider them as two stellar clumps within a single dark matter halo, two arguments suggest a halo mass $\log(M_h/M_\odot) \approx 10.3$ to 10.4. The derived stellar mass for the combined system is $\log(M_\ast/M_\odot) = 8.16$.  The stellar mass -- halo mass relation at $z \approx 10$ derived by \citet{Behroozi2019} via abundance matching would then imply $\log(M_h/M_\odot) \approx 10.3$.  Alternatively, if the observed $\Delta v$ is interpreted as relative motions of two components with orbital circular velocities $v_c = 110 \pm 50$~km~s$^{-1}$ (likely a lower limit given projection effects), virial relations for a dark matter halo collapsed at $z = 9.8$ imply $\log (M_h/M_\odot) = 10.4_{-0.8}^{+0.5}$ \citep{Barkana_Loeb_2001}.  Recent semi-analytic models that reproduce the observationally estimated stellar mass function at $z\sim 9$ \citep{yung2024a} give very similar results.  However, the virial radius for such a halo would be $r_\mathrm{vir} = 8.6$~kpc, nearly 10$\times$ the observed transverse separation. The peak rotation velocity for a \citet[][NFW]{NFW1997} dark matter halo occurs at a radius of about twice the NFW scale radius, which is in the range 4.8 to 6.7 kpc given the typical halo NFW concentration of $c_{\rm NFW}\sim 3$ at $z\sim 10$ \citep{yung2024b}, again many times the projected separation. This suggests that the observed $\Delta v$ indicates motion along the line of sight, more like an encounter of two objects than rotation within a common halo. Furthermore, the projected separation of the two components (0.93 kpc) is much larger than the sizes of most galaxies at $z \approx 10$ \citep[e.g.,][]{Morishita2024}, and the luminosities and derived stellar masses of the two components are large for sub-galactic stellar clusters. Taken together, these arguments suggest that 52185 and 300001 are two galaxies encountering one another and perhaps likely to merge.  However, the imprecision of the NIRSpec PRISM redshifts limits this analysis. Integral field spectroscopy at higher spectral resolution with the NIRSpec gratings would clarify the interpretation.

\subsection{The nature of blue UV slopes}
\label{sec:cosmos_beta}
The steepness of the UV slope is defined by both the stellar and nebular continua. AGN emission can also affect the continuum slope, and is generally redder than the measured $\beta$ values for these COSMOS galaxies, but we find no clear indication of AGN activity from the spectra of these sources.
Younger stellar age, lower metallicity and lower dust attenuation all steepen the UV slope of the stellar continuum. A galaxy with a purely stellar spectrum with extremely young ages, extremely low metallicity and a lack of dust attenuation can have a UV slope of $\beta\lesssim-3$ and detecting this has been proposed as a method for detecting the presence of Population III stars \citep{schaerer2002}. However, star-forming galaxies typically contain ionized gas which produces both nebular emission lines as well as continuum emission. Nebular continuum emission can redden the UV slope of a galaxy, and if one assumes a maximum contribution from nebular emission and a standard slope and cutoff to the high-mass IMF, the bluest allowed UV slope is $\beta\simeq-2.6$ with different models suggesting the minimum is in the range $\beta\simeq-2.4$ to $\beta\simeq-2.6$ \citep{stanway2016,topping2022,cullen2023}. In Fig.~\ref{fig:beta_relations} we show this minimum allowed UV slope by the shaded blue region. 
Therefore, an observed UV slope steeper than $\beta \simeq-2.6$ likely indicates the galaxy has very low or zero dust attenuation as well as a low contribution from nebular continuum, which in turn may imply a non-negligible Lyman continuum escape fraction.

The observed UV slopes for COSMOS-52815 and COSMOS-300001 are both significantly steeper/bluer than is typical for bright galaxies ($M_{\rm UV}\lesssim-20$) at $z\simeq10$ as observed from large photometric samples \citep[e.g.,][]{morales2023,cullen2023} as well as the current sample of other spectroscopically confirmed galaxies at $z>9$ \citep[e.g.,][]{dottorini2024}. UV slopes this blue have been observed at $z\simeq10$ but for galaxies $\sim1-2\, \rm mag$ fainter \citep[e.g.,][]{yanagisawa2024}, as well as at lower redshift \citep{jiang2020,bolamperti2023}. Both sources are consistent with little-to-no dust attenuation as evidenced by their steep UV slopes. This is also consistent with the results of SED fitting with \textsc{bagpipes} where we obtain $A_{\rm V} = 0.07^{+0.05}_{-0.03}$ for COSMOS-52185 and $A_{\rm V} = 0.02^{+0.03}_{-0.01}$ for COSMOS-300001 (assuming a fixed dust-attenuation law). This is consistent with the expectations of recent theoretical models. \citet{ferrara2023,ferrara2024} suggest that for bright galaxies at $z>10$, there is dust production but then it is ejected or destroyed in galaxies where $\rm sSFR>30\, Gyr^{-1}$ causing an observed lack of dust attenuation (the COSMOS galaxies have $\rm sSFR\sim90\, Gyr^{-1}$). \citet{donnan2025} show that the galaxy population transitions to being largely dust-free by $z\simeq12$ because galaxies have stellar masses too low to contain significant levels of dust even if the $A_{1500}-M_{\star}$ relation is unchanged since $z=2$. COSMOS-52815 and COSMOS-300001 are consistent with this relation as they have low stellar masses $\log(M_{\star}/\rm M_{\odot})\simeq7.9$ showing that even at $z=2$ (where the fraction of obscured star-formation peaks), these galaxies would have little dust attenuation.

Given the detection of the H$\beta$ and \oii\, emission lines, we use the O2 strong-line metallicity calibration from \citet{sanders2025} to provide an estimate of the metallicity for the COSMOS pair. This yields low metallicity measurements of $Z=0.15\pm0.07 \rm \, Z_{\odot}$ and $Z=0.06\pm0.17 \rm \, Z_{\odot}$ for COSMOS-300001 and COSMOS-52185 respectively.

\subsection{Implications for cosmic reionization}
COSMOS-300001 has one of the bluest UV slopes at $z>9$ observed with $\beta=-2.87\pm0.15$ and as shown in the right-hand panel of Fig.~\ref{fig:beta_relations}, it is the bluest UV slope observed to date in a bright galaxy ($M_{\rm UV}\lesssim-20$) at $z>9$. As we discussed in Section~\ref{sec:cosmos_beta}, this suggests that COSMOS-300001 cannot have significant dust attenuation. It is also bluer than the limit of $\beta \simeq-2.6$ for a galaxy with a full contribution from stellar and nebular continuum even with no dust attenuation. Therefore, this suggests a weakened contribution from the nebular continuum. This is also evidenced by the weaker nebular emission lines observed in COSMOS-300001 compared to COSMOS-52185 (e.g., H$\beta$), although their presence in 300001 does signpost the presence of ionized gas ($\mathrm{EW_0} = 104\pm32$\AA\ for H$\beta$ is hardly negligible). The blue UV slope could therefore indicate the presence of strong outflows which create channels in the ISM to allow a significant fraction of ionizing radiation to escape \citep{ferrara2025}.

Studies of low-redshift star forming galaxies have demonstrated a strong correlation between Lyman continuum escape fraction and UV spectral slope \citep{chisholm2022}. Applying this same correlation at high redshift, the very steep UV slopes of the COSMOS pair galaxies imply significant ionizing escape fractions: $f_{\rm esc} = 0.41\pm0.20$ for COSMOS-300001 and $f_{\rm esc}=0.12\pm0.04$ for COSMOS-52815. Weaker EW(H$\beta$) for 300001 compared to 52185 would also be qualitatively consistent with a larger $f_{\rm esc}$. We also derive $f_{\rm esc}$ by fitting the photometry and spectrum (including emission lines) using \textsc{bagpipes} where we allow $f_{\rm esc}$ to be a free parameter following the methodology of \citet{papovich2025}. With this method, we find lower escape fractions than the values predicted by the $\beta-f_{\rm esc}$ relationship (consistent with the overall findings of \citealp{papovich2025}), but a similar trend, with larger escape fractions predicted (or at least allowed) for COSMOS-300001 ($f_{\rm esc}=0.02^{+0.14}_{-0.02}$) compared to COSMOS-52815 ($f_{\rm esc}=0.01^{+0.04}_{-0.01}$). For COSMOS-300001, the 84\% confidence range on $f_{\rm esc}$ extends to $\sim 3\times$ the corresponding range for COSMOS-52185, and the tail of the probability distribution extends as high as 30\%, which is very unusual in the sample analyzed by \citet{papovich2025}. Although this highlights that the absolute value of $f_{\rm esc}$ is highly method-dependent, the steep UV slope of COSMOS-300001 does suggest higher $f_{\rm esc}$ than is typical for galaxies at $z>9$. 

Given the close presence of two UV luminous galaxies together during the epoch of reionization, we explore the possibility of the total ionizing output carving an ionized bubble. This is of particular interest given the lack of Ly$\alpha$ detection in either galaxy. Following the method of \citet{napolitano2024b}, we are able to place strong upper limits on the $3\sigma$ rest-frame equivalent width of Ly$\alpha$ of $<15\, \mathrm{\AA}$ and $<11\, \mathrm{\AA}$ for COSMOS-300001 and COSMOS-52185 respectively. Using the measured $M_{\rm UV}$ and $\beta$ to infer the rate of ionizing photon escape using the \citealp{chisholm2022} method, and the stellar age $t_{\star}$, we determine $R_{\mathrm{ion}}=0.2\pm0.1\, \mathrm{Mpc}$ following the method of \citet{napolitano2024b}. This is only $\sim10\%$ the expected size of an ionized bubble that would allow Ly$\alpha$ to escape (and would be even smaller using $f_{\mathrm{esc}}$ derived from SED fitting), consistent with the lack of a detection of Ly$\alpha$ in this galaxy \citep[e.g.][]{lu2024}.

Alternatively, we can use the ratio of the luminosity of H$\beta$ to the UV luminosity to estimate the ionizing photon production efficiency, $\xi_{\rm ion}$. Using the relation from \citet{emami2020} assuming no dust reddening and $f_{\rm esc}=0$ (chosen given the uncertainty on the exact value described above) we derive $\xi_{\rm ion}=25.0\pm0.1$ and $\xi_{\rm ion}=25.5\pm0.1$ for COSMOS-300001 and COSMOS-52185 respectively. These values are consistent with typical ionizing photon production efficiencies at $z\geq6$ \citep[e.g.][]{simmonds2024b,begley2024,llerena2025}.

\section{A red, dusty star-burst at $\lowercase{z}=9.94$?} \label{sec:discussion2}
Unlike COSMOS-300001 and COSMOS-52185, EGS-25297 has an extremely red slope for a galaxy at $z\simeq10$ with $\beta=-1.51\pm0.08$. As shown in Fig.~\ref{fig:beta_relations}, this is one of the reddest UV slopes detected at $z>9$ to date. Although similar in its (observed) UV luminosity and redshift to the COSMOS pair, this red UV slope suggests there could be a strong nebular continuum, a red AGN continuum or simply dust-reddened star-formation in EGS-25297. In this section we evaluate each of these possibilities. 

\subsection{Nebular continuum}
\begin{figure}[ht]
\centering
\includegraphics[width=\columnwidth] {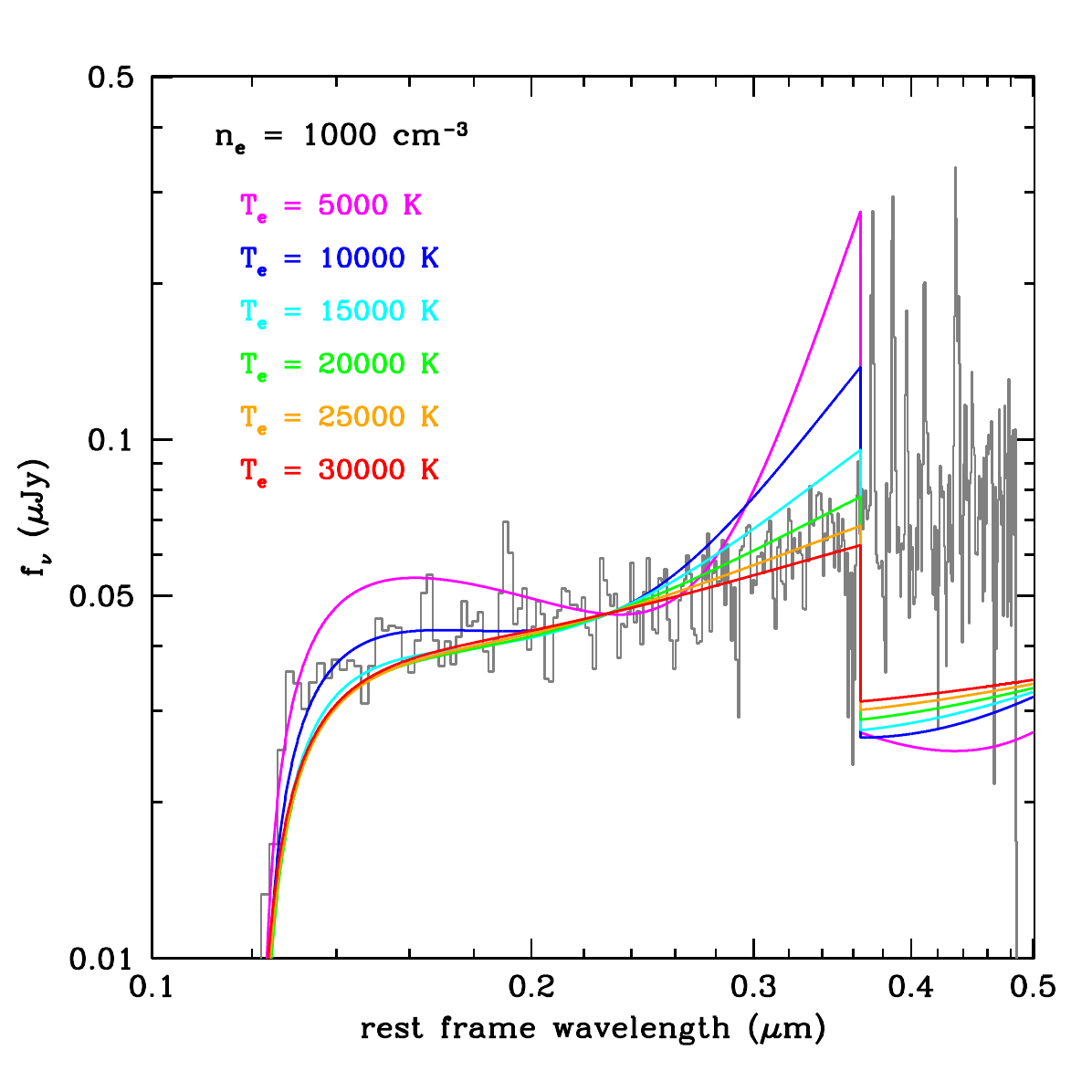}
\caption{\textsc{PyNeb} models for pure hydrogen nebular continuum at fixed electron density $n_e = 1000$~cm$^{-3}$ and varying temperature $T_e$, normalized to match the spectrum of EGS-25297 at 2350\AA\ in the rest frame.  All models predict large Balmer ``jumps'' at 3646\AA\ that are not observed for EGS-25297.  Models with higher $n_e$ and large $T_e$ (not shown here) predict UV continuum slopes that are too red compared to the observations. The 2-photon continuum component of the high temperature nebular models does not match the shape of the observed continuum at $\lambda < 1400$\AA.}
\label{fig:nebular_continuum}
\end{figure}

Strong nebular continuum emission can exhibit a redder UV spectral slope than that of a young stellar population with low reddening.  However, strong UV nebular continuum emission is accompanied by a large Balmer ``jump'' -- a sharp increase in flux density at wavelengths shorter than the 3646\AA\ Balmer limit \citep{cameron2024}.  Nebular continuum models computed with \textsc{PyNeb} \citep{Luridiana2015} at high electron temperatures $T_e > 20000$~K and low densities $n_e \lesssim 1000$~cm$^{-3}$ can roughly match the UV continuum shape and slope observed for EGS-25297, but predict very large Balmer jumps that are not observed for this galaxy (Fig.~\ref{fig:nebular_continuum}). They also predict a more gradual ``roll-off'' at $\lambda_{\rm rest}\lesssim1400\, \rm \AA$ than is observed, driven by two-photon continuum emission. A further possibility is the presence of an older stellar population which creates a Balmer break that cancels the Balmer jump from the nebular continuum. However, this is unlikely due to the requirement for strong UV emission from young O and B type stars to ionize and heat gas to high temperatures to generate nebular continuum, which would in turn outshine and dilute the Balmer break produced by the older stellar population.

\subsection{Emission line properties}
\label{sec:25297_emission_lines}
Given the significant ($>3\sigma$) detection of the electron-temperature sensitive \oiiiA\, emission line in EGS-25297, we can estimate the oxygen abundance via the ``direct-$T_e$'' method. 
As we do not resolve the density sensitive \oii\, doublet and do not cover the \sii\, doublet, we employ a forward modeling approach using {\sc PyNeb} (e.g., \citealp{cullen2025, scholte2025}).
We construct a model of five free parameters: electron temperature ($T_e\,{\rm [O\, III]}$), electron density ($n_e$), the abundance of singly ionized oxygen ($\log (\rm O^{+}/{H^{+}})$), the abundance of doubly ionized oxygen ($\log (\rm O^{2+}/{H^{+}})$), and the $V$-band dust attenuation ($A_V$).
We then explore the parameter space using the {\sc dynesty} nested sampling code \citep{speagle2020, koposov2023}.
For each parameter set we calculate the emissivities of the observed lines, H$\delta$, H$\gamma$, H$\beta$, \oiiiA, \oiiida, and \oiiidb, and apply reddening using the model $A_V$ (assuming a \citealp{cardelli1989} attenuation prescription).
We then compare ratios of these emission lines to the observed (i.e., not dust corrected) emission line ratios to determine the best fitting model.
Using forward modeling is advantageous compared the traditional multi-step approach as we are able to marginalize over electron density, and we can fit the observed line flux ratios by including the $V$-band attenuation as a free parameter.
This approach has also been shown to yield consistent results with the standard methodology \citep[e.g.,][]{arellano-córdova2025, scholte2025, stanton2025}. Applying the more traditional methodology for EGS-25297 does not change our resulting metallicity.

Our best fitting model yields an electron temperature of $T_e\,[\rm O\,III]=17200\pm1900\,{\rm K}$ and a total oxygen abundance of $12 + \log \rm (O/H)=7.70\pm0.13$, which corresponds to $Z\simeq0.1\,\rm Z_{\odot}$ assuming a solar metallicity of $12+\log(\rm O/H) = 8.69$ \citep{asplund2021}
This is fully consistent with the recent analysis by \citet{pollock2025}, who estimated the direct-method oxygen abundances of a sample of $z\sim10$ galaxies, finding an electron temperature and oxygen abundance for EGS-25297 of $T_e = 17000\pm1200\,{\rm K}$ and $12+\log (\rm O/H) = 7.69 \pm 0.08$ respectively.
Interestingly, our result is also fully consistent with measurements using the strong-line calibrations of $\widehat{RNe}$ ($12+\log(\rm O/H) = 7.67\pm0.47$) and Ne3O2 from \citet{scholte2025} as well as the value derived in SED fitting (Table~\ref{tab:SED_properties}).
This places EGS-25297 in line with the expected metallicity for galaxies of equivalent stellar mass on the mass-metallicity relationship at $z=10$ \citep{nakajima2023,curti2024}. 
Our best fit model also finds $A_V = 0.81^{+1.01}_{-0.72}$; whilst not tightly constrained, this estimate of $A_V$ is reassuringly consistent with the $A_V$ derived from fitting with {\sc bagpipes} of $0.92^{+0.42}_{-0.27}$.

For EGS-25297, we also obtain marginal ($\sim2\sigma$) detections of the \ciii\, and blended \heii\ + \oiiiuv, rest-frame UV emission lines. The rest-frame equivalent width of the \ciii\, line has been observed to increase on average with increasing redshift \citep[e.g., $\mathrm{EW}_{0}=1.84z-5.38$;][]{robertsborsani2024}. We measure a rest-frame equivalent width for EGS-25297 of $22.6\pm10.5\, \rm \AA$ and is therefore consistent with this relationship within $1\sigma$. Although we fail to detect any rest-frame UV emission lines for the COSMOS pair, their upper limits exclude the possibility of extreme emission (e.g., $\rm EW_{0}>20\, \AA$) for \ciii, \heii\, and \civ\, unlike what has been observed in some other galaxies at $z>9$ \citep[e.g., GHZ2;][]{castellano2024}. We can therefore exclude the possibility of extreme emission for any UV emission lines in these three sources. 

\subsection{Does EGS-25297 host an AGN?}
\label{sec:25297_agn}
\textit{JWST} has revealed a significant population of broad-line AGN that have been named ``Little Red Dots'' \citep[LRD;][]{matthee2024}. Since this initial discovery, there have been a number of studies that have selected LRDs based on photometric colors \citep{kocevski2025,perezgonzale2024} due to their blue rest-frame UV emission and red rest-frame optical emission. Recently, it has been demonstrated that a number of LRDs with confirmed broad-line detections, exhibit similarly red UV slopes to that of EGS-25297 \citep{greene2024}. We therefore discuss the possibility of an AGN in EGS-25297 and review the available evidence.

Although this galaxy had been primarily selected as a $z>9$ galaxy candidate \citep[e.g.,][]{mcleod2023, finkelstein2024} it had also been selected as an LRD candidate by \citet{kocevski2025}, based on rest-frame optical and UV slopes, requiring $\beta_{\rm UV}<-0.37$ and $\beta_{\rm opt}>0$. EGS-25297 met this criteria due to the previous photometry-based UV and optical slopes (and photometric redshift $z_{\rm phot}=10.21$) of $\beta_{\rm UV}=-1.47 \pm 0.15$ and $\beta_{\rm opt}=-0.02 \pm 0.24$. However, now that we have a NIRSpec spectrum of this galaxy, it is apparent that the red rest-frame optical slope is partly driven by the presence of strong emission lines which dominate the flux observed in the F410M and F444W filters. Using the NIRSpec spectrum (and spectroscopic redshift) and masking all detected emission lines, we now measure a rest-frame optical slope of $\beta_{\rm opt}=-0.65\pm 0.35$. As this removes the impact of emission lines and therefore purely traces the underlying continuum, we find a significantly bluer rest-frame optical slope such that this object would fail LRD selection based on its continuum SED alone.

The low spectral resolution of the NIRSpec PRISM makes it difficult to determine if there are broad components to the Balmer emission lines.  No strong broad emission with FWHM~$> 1500$~km~s$^{-1}$ is seen for $H\gamma$ or H$\beta$, but careful attention is required to decompose narrow and broad  components.  New $R\simeq1000$ NIRSpec G395M observations (GO 5718, PI: D.\ Kocevski) should be able to definitively determine the presence or absence of broad line emission in this galaxy. 

The detection of UV emission lines can also be used to determine the presence of AGN activity as they are sensitive to the hardness of ionizing radiation. Because of this a number of rest-frame UV line ratio diagnostics have been developed \citep[e.g.,][]{feltre2016,nakajima2018,hirschmann2019}. However, due to the lack of detection of these key emission lines, this method of determining the presence of an AGN is unavailable. However given that the 2$\sigma$ upper limits on the rest-frame UV emission lines are $\lesssim15\, \rm \AA$, this suggests that extreme emission from these UV lines which would be expected in the case of an AGN is absent in EGS-25297. 

Given the lack of detected UV emission lines, we can in principle use the rest-frame optical AGN diagnostic diagram from \citet{mazzolari2024} to test if the measured line-ratios of \oiiiA/H$\gamma$ to \neiii/\oii\, can confirm/rule-out the presence of an AGN. However, this diagnostic can only confirm the presence of an AGN particularly in the case of a strong \oiiiA/H$\gamma$ ratio. For EGS-25297, we find that it is consistent with AGN or star-formation by this diagnostic and therefore we cannot use this to exclude the possibility of an AGN within EGS-25297.

\subsection{A dust-reddened star-forming galaxy at $z=10$}
\label{sec:25297_beta}
Given that it is unlikely that the red UV slope is explained by strong nebular continuum and with a lack of evidence for AGN activity, the final obvious option is that EGS-25297 is a dust-reddened star-forming galaxy.
Indeed our best-fitting SED model from \textsc{bagpipes} suggests a V-band attenuation of $A_{\rm V}=0.92^{+0.42}_{-0.27}$~mag. Given that this galaxy is observed to be already luminous in the UV ($M_{\rm UV}=-20$), this significant level of dust attenuation suggests that EGS-25297 has an extremely high intrinsic UV luminosity. Assuming a \citet{calzetti2000} dust-attenuation law, we predict a UV attenuation of $A_{1500}\simeq2.35^{+1.07}_{-0.69}$~mag which implies an intrinsic UV luminosity of $M_{\mathrm{UV,corr}}\simeq -22.4^{+1.07}_{-0.69}$.

To place this into context, we compare EGS-25297 to the most UV-luminous, spectroscopically confirmed galaxies known at $z > 9$.  Firstly, GN-z11 has an observed UV magnitude of $M_{\rm UV}=-21.5$ at $z=10.6$ \citep{bunker2023b}. However unlike EGS-25297, GNz11 has a significantly bluer UV slope of $\beta=-2.26\pm0.10$. From SED fitting, \citet{bunker2023b} report a V-band attenuation of $A_{\rm V}=0.17$. Assuming a \citet{calzetti2000} dust-attenuation law, this implies an intrinsic UV magnitude of $M_{\mathrm{UV,corr}}\simeq-21.93$. Similarly for Gz9p3, a $M_{\rm UV}=-21.66$ galaxy at $z=9.31$, \citet{boyett2024} measure a UV slope of $\beta=-2.23\pm0.04$ and from SED fitting find a V-band attenuation of $A_{\rm V}=0.2$ implying an intrinsic UV magnitude of $M_{\mathrm{UV,corr}}=-22.17$. Finally, for GHZ2 \citet{castellano2024} report minimal dust attenuation ($A_{\rm V}=0.04$) and therefore this galaxy at $z=12.34$ has an intrinsic UV magnitude of $M_{\mathrm{UV}, corr}=-20.63$. Therefore, a standard correction for starburst reddening would imply that EGS-25297 is the intrinsically most UV luminous galaxy yet identified and spectroscopically confirmed at $z>9$ (Fig.~\ref{fig:Muv_z}). 

EGS-25297 also displays a damped Lyman-$\alpha$ break suggesting the presence of a significant local reservoir of neutral hydrogen with  $\log(n_{\mathrm{HI}}/\mathrm{cm}^{-2}) = 21.9\pm0.3$ (see Section~\ref{sec:ly_break}) and a high covering factor over the galaxy's UV continuum emission. 
It seems plausible that a large $n_{\mathrm{HI}}$ could be associated with greater dust reddening compared, e.g., to the low $n_{\mathrm{HI}}$ and low reddening for the COSMOS pair.  That said, the physical nature and particularly the spatial location of the gas responsible for the Ly~$\alpha$ damping observed for many $z > 5$ galaxies is still not well understood \citep[e.g.,][]{heintz2024, umeda2024}.   While the ionized gas in EGS-25297 has $Z \simeq 0.1 Z_\odot$ (Section~\ref{sec:25297_emission_lines}), the high neutral column density could indicate accretion of relatively pristine intergalactic HI with little connection to the observed reddening.

\begin{figure}[ht]
\centering
\includegraphics[width=\linewidth]
{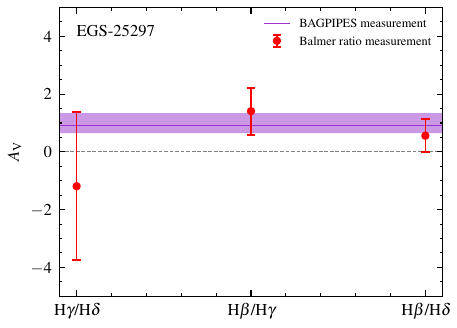}
\caption{A comparison of the $A_{\rm V}$ derived from the Balmer line ratios of H$\gamma$/H$\delta$, H$\beta$/H$\gamma$ and H$\beta$/H$\delta$ for EGS-25297 (assuming a \citealp{cardelli1989} attenuation curve), to the value derived from SED fitting with \textsc{bagpipes}.}
\label{fig:balmer_Av}
\end{figure}

Given the successful detection of \Hg\,, \Hd\, and \Hb\, we can also measure the dust attenuation using the ratio of these Balmer recombination lines. Assuming Case B recombination with a temperature of $20000\, \rm K$, the intrinsic ratios are expected to be H$\gamma$/H$\delta=1.79$, H$\beta$/H$\gamma=2.06$ and H$\beta$/H$\delta=3.69$ \citep{osterbrock1989}. In the case of dust reddening, we expect these ratios to exceed their intrinsic values. For EGS-25297, we measure H$\gamma$/H$\delta=1.63\pm0.30$ which is consistent with the intrinsic ratio and therefore it appears that there is an inconsistency in the implied dust attenuation from SED fitting (driven by the red UV slope) compared to the H$\gamma$/H$\delta$ ratio. However, the uncertainty on the measured H$\gamma$/H$\delta$ ratio translates to an uncertainty on the V-band attenuation of $\Delta A_{\rm V}\simeq2$ and is therefore entirely consistent with the $A_{\rm V}$ derived from SED fitting. Indeed, the measured ratios of both H$\beta$/H$\gamma=2.61\pm0.35$ and H$\beta$/H$\delta=4.24\pm0.58$ suggest some level of dust attenuation. Assuming a \citet{cardelli1989} nebular attenuation law, we determine the $A_{\rm V}$ implied from each Balmer line ratio and compare to the value derived from \textsc{bagpipes}. This is shown in Fig.~\ref{fig:balmer_Av} demonstrating consistency between the measurements. This calculation assumes the same level of dust reddening for the gas and stars within the galaxy.

We also use the emission line flux of H$\beta$ to calculate a star-formation rate (SFR). After dust-correcting the flux, there is a ratio of H$\alpha$/H$\beta=2.69$ \citep{osterbrock1989} which enables us to determine the H$\alpha$ flux. Using the SFR calibration from \citet{reddy2018} which assumes a \citet{chabrier2003} IMF, we then find that the H$\beta$ flux implies $\mathrm{SFR}=50^{+28}_{-12}\, \rm M_{\odot}/yr$. In Table~\ref{tab:SED_properties} we reported the $\rm SFR_{10}$ for EGS-25297 using the DPL SFH of $117^{+21}_{-18}\, \rm M_{\odot}/yr$. However, this value varies depending on the chosen form of the SFH (further discussed in Section~\ref{sec:25297_SFH}) with a lower value of $\mathrm{SFR_{10}}=76^{+13}_{-14}\, \rm M_{\odot}/yr$ for the non-parametric SFH. Therefore although quite uncertain, this is broadly consistent with the measured SFR over the last $10\, \rm Myr$ from the SFH derived from SED fitting, providing further evidence that this source is a dust-reddened galaxy undergoing a strong episode of star-formation.

\subsection{Constraints on the star-formation history}
\label{sec:25297_SFH}
At $z\sim2$ there has been an observed relationship between stellar mass and the UV dust attenuation, $A_{1500}$, showing that more massive galaxies have increased dust attenuation \citep{garn2010,mclure2018}. For EGS-25297, (assuming a \citealp{calzetti2000} attenuation law) we determine $A_{1500}\simeq2.3$ which would typically correspond to stellar masses of $\log(M_{\star}/\rm M_{\odot})\simeq10$. This is significantly in excess of the stellar mass we derive using a DPL SFH from \textsc{bagpipes} ($\log(M_{\star}/\rm M_{\odot})=8.95^{+0.04}_{-0.05}$). However given we only have observations out to $\lambda_{\rm rest}\simeq4900\, \rm \AA$, it entirely possible to have a SFH that produces a near-identical SED to match the observations but yields a significantly greater mass. This effect has been observed in galaxies previously and is often known as ``outshining'' \citep{papovich2001,conroy2010,Pforr2012}. Indeed, in the right-hand panel of Fig.~\ref{fig:SFH} we show that assuming a ``continuity-bursty'' \citep{tacchella2022} non-parametric SFH, we obtain a significantly greater stellar mass of $\log(M_{\star}/\rm M_{\odot})=9.60^{+0.09}_{-0.12}$ due to a previous episode of star-formation. In this SFH, this star-formation occurs at $z\simeq20$ which would imply $M_{\rm UV}\simeq-21.8$ at this redshift. However, the exact form and timescale of this possible past star-formation remains completely uncertain and it is possible to fit the data with a variety of SFH's. Therefore, although past star-formation allows a more reasonable stellar mass to compare to the $A_{1500}-M_{\star}$ relation at lower redshift, given the current available data it is impossible to confirm or deny this and this emphasizes the importance of longer-wavelength observations to provide a more accurate stellar mass. If this increased stellar mass is confirmed, it would be one of the most massive galaxies at $z>9$, exceeding that of Gz9p3 \citep[$\log(M_{\star}/ \rm M_{\odot})=9.2$;][]{boyett2024} and GHz2 \citep[$\log(M_{\star}/ \rm M_{\odot})=9.1$;][]{castellano2024}. 

Assuming the H$\beta$ derived value for $\rm SFR_{10}$ and the DPL SFH stellar mass of $\log(M_{\star}/\rm M_{\odot})=8.95$, we obtain $\rm sSFR\simeq60\, Gyr^{-1}$ for EGS-25297 which would suggest there should be dust ejection according to the \citet{ferrara2024} model. This is clearly inconsistent with the observed dust-attenuation in this galaxy potentially suggesting there are galaxies at $z>9$ that can achieve high $\rm sSFR$ without ejecting their dust. Alternatively, this inconsistency can be resolved if we assume the greater stellar mass obtained from the non-parametric SFH ($\log(M_{\star}/\rm M_{\odot})=9.6$) yielding $\rm sSFR\simeq12\, Gyr^{-1}$ which is significantly below the threshold for dust ejection in the \citet{ferrara2024} model ($\rm sSFR=30\, Gyr^{-1}$).

Given the presence of dust in this galaxy ($A_{\rm V}\simeq0.92$), this may suggest a requirement for a previous episode of star-formation in order to assemble sufficient dust by the time of observation. Typically in galaxies, the majority of dust assembles through grain growth \citep{draine2009} and this can happen on short timescales ($<1\,\rm Myr$). However, this typically only occurs in galaxies where the gas-phase metallicity is $>0.2\, \rm Z_{\odot}$ \citep{galliano2021} and is therefore unlikely to occur in EGS-25297. Type II supernovae can produce dust in low-metallicity galaxies and are often assumed to dominate the dust production in high-$z$ galaxies \citep[e.g.,][]{todini2001,lesniewska2019} and can produce dust on very short timescales. Therefore, it is perhaps not necessary to require significant older star-formation to produce an $A_{\rm V}\simeq0.9$ by the time of observation.

Finally, another explanation for the red UV slope 
could be the presence of a significantly older stellar population reddening the UV slope instead of dust with a recent upturn in star-formation driving the observed emission lines. Sources with such properties have been observed at $z\sim4-7$ \citep[e.g.,][]{strait2023,looser2024}. However, this would require significant star-formation at $z>20$ and the strong equivalent widths of the observed emission lines require a dominant, young, actively star-forming stellar population which would still dominate the UV continuum (and therefore require dust to create the observed red UV slope) making this scenario also unlikely.

Overall, although the data cannot exclude the possibility of significant past star-formation (either producing significant dust and/or older stellar populations), the data is entirely consistent with simply a young, dust-reddened star-forming galaxy.

\section{Conclusions} \label{sec:conclusion}
In this paper we report \textit{JWST}/NIRSpec PRISM spectroscopic observations of two galaxies at $z=9.8$ that have very blue UV slopes and one galaxy at $z=9.94$ with a red UV slope. We conclude the following:
\begin{itemize}
    \item We measure a spectroscopic redshift based on the presence of a number of emission lines for 3 luminous ($M_{\rm UV}\lesssim-20$) galaxies at $z\sim10$. This includes 2 blue galaxies (COSMOS-300001, COSMOS-52185) selected from NIRCam imaging in PRIMER COSMOS with a spectroscopic redshift of $z=9.800\pm0.003$ and $z=9.808\pm0.002$, and a red galaxy selected from NIRCam imaging in CEERS EGS (EGS-25297) at $z=9.942\pm0.002$. 
    \item We measure significantly bluer UV slopes for the COSMOS galaxies than is typical for luminous $z=10$ galaxies ($\beta\sim-2.2$) with $\beta=-2.87\pm0.15$ and $\beta=-2.43\pm0.10$. From spectrophotometric modeling with \textsc{bagpipes}, this implies young mass-weighted ages $t_{\star}\sim2\, \rm Myr$ and a lack of dust attenuation ($A_{\rm V}<0.1$).
    \item The COSMOS galaxies are separated by only 0.2-arcsec and a velocity difference of $\Delta v=220\pm100\, \rm km\, s^{-1}$ potentially indicating they are actively merging. A merger can drive the burst of active star-formation, with the UV flux dominated by extremely young stars and a lack of dust creating an extremely blue UV slope.
    \item For EGS-25297, we measure a red UV slope of $\beta=-1.51\pm0.08$ which makes it one of the reddest galaxies detected at $z>9$ to date. The spectrum disfavors the possibility of a strong contribution from the nebular continuum due to the lack of a Balmer jump, and we find no evidence to support the presence of an AGN, although this cannot be definitively excluded with the current data. Instead, the red UV slope is likely due to significant dust attenuation ($A_{\rm V}\simeq0.9$), potentially suggesting some level of dust obscured star-formation in this luminous galaxy at $z=9.94$. 
\end{itemize}

This study demonstrates that spectroscopy with \textit{JWST}/NIRSpec has the ability to not only confirm the redshift of galaxies with extreme UV slopes at $z>9$, but also determine the physical drivers of these UV slopes. The complete sample of galaxies at $z>9$ observed with CAPERS will provide valuable statistics on the diversity of UV slopes, and follow-up higher resolution NIRSpec observations as well as those with other instruments/facilities, can aid our understanding of the diverse galaxy population at $z>9$.\\

\section*{Acknowledgements}
\noindent We thank Eduardo Ba\~{n}ados for providing access to the NIRCam F090W data for the CEERS EGS field (GO-2234). F. Cullen and T. M. Stanton acknowledge support from a UKRI Frontier Research Guarantee Grant (PI Cullen; grant reference EP/X021025/1). 

This work is based on observations made with the NASA/ESA/CSA \textit{James Webb Space Telescope}, obtained at the Space Telescope Science Institute, which is operated by the Association of Universities for Research in Astronomy, Incorporated, under NASA contract NAS5-03127. Support for program number GO-6368 was provided through a grant from the STScI under NASA contract NAS5-03127. The data were obtained from the Mikulski Archive for Space Telescopes (MAST) at the Space Telescope Science Institute. 
These observations are associated with program \#6368, and can be accessed via \dataset[doi: 10.17909/0q3p-sp24]\\

Some of the data products presented herein were retrieved from the Dawn JWST Archive (DJA). DJA is an initiative of the Cosmic Dawn Center (DAWN), which is funded by the Danish National Research Foundation under grant DNRF140.


%

\vspace{5mm}
\facilities{JWST, HST}


\software{astropy \citep{2013A&A...558A..33A,2018AJ....156..123A},  
          \textsc{bagpipes} \citep{carnall2018}, \textsc{galfit} \citep{peng2002}
          }



\appendix
\section{DJA extended wavelength reduction}
\label{sec:apdx_A}
\begin{figure*}[ht]
\centering
\includegraphics[width=\linewidth]
{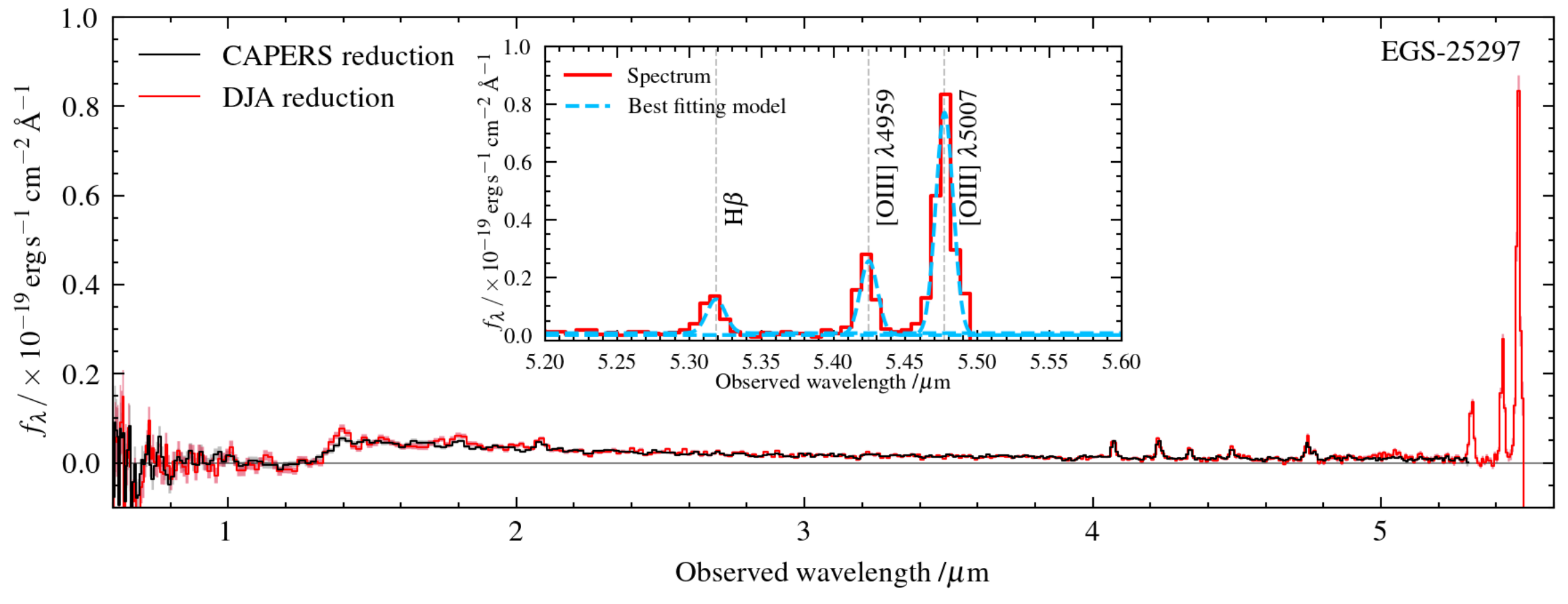}
\caption{The 1-D spectra of EGS-25297 from the CAPERS reduction (black) and the Dawn JWST Archive (DJA; red) which extends the spectrum out to $\lambda_{\rm observed}=5.5\, \mu$m. The inset panel shows the best fitting Gaussian models (blue) to the emission lines at $\lambda>5.3\, \mu$m.}
\label{fig:extended_spec}
\end{figure*}

Due to the spectroscopic redshift of EGS-25297 being $z=9.942$, the H$\beta$ emission line is redshifted beyond the default maximum wavelength of the NIRSpec reduction pipeline of $\lambda=5.3\, \mu$m. As mentioned in Section~\ref{sec:emission_lines}, we therefore use the \textsc{MSAexp} \citep{Brammer2023} reduction from the Dawn JWST Archive \citep[DJA v4;][]{valentino2025,pollock2025}, which extends the wavelength range of the NIRSpec PRISM extraction to 5.5$\,\mu$m, to measure the line fluxes of H$\beta$, \oiiida\, and \oiiidb. We follow the same procedure as all other line flux measurements in this work, as described in Section~\ref{sec:emission_lines}. In Fig.~\ref{fig:extended_spec} we show a comparison of our primary reduction and the extended wavelength reduction from the DJA with the results of our line fits to H$\beta$, \oiiida\, and \oiiidb\, shown in the inset panel.  At $\lambda < 5.3\,\mu$m, the CAPERS and DJA reductions match one another closely.


\bibliography{extreme_UV_slopes_z10}{}
\bibliographystyle{aasjournal}



\end{document}